\documentclass[prb, reprint, groupedaddress]{revtex4-1} 
\usepackage{graphicx}
\usepackage{amsmath}
\usepackage{amssymb}
\usepackage{soul}

\begin{document}
\title{Scanning gate microscopy of magnetic focusing in graphene devices:
       quantum vs.~classical simulation}
\author{M.~D.~Petrovi\'c}
\author{S.~P.~Milovanovi\'c}
\author{F.~M.~Peeters}
\affiliation{Department of Physics, University of Antwerp, \\
             Groenenborgerlaan 171, B-2020 Antwerp, Belgium}

\begin{abstract}

We compare classical versus quantum electron transport in recently
investigated magnetic focusing devices [S. Bhandari et al., Nano Lett.~16,
1690 (2016)] exposed to the perturbing potential of a scanning gate microscope
(SGM). Using the Landauer-B{\"u}ttiker formalism for a multi-terminal device,
we calculate resistance maps that are obtained as the SGM tip is scanned over
the sample. There are three unique regimes in which the scanning tip can
operate (focusing, repelling, and mixed regime) which are investigated. Tip
interacts mostly with electrons with cyclotron trajectories passing directly
underneath it, leaving a trail of modified current density behind it. Other
(indirect) trajectories become relevant when the tip is placed near the edges
of the sample, and current is scattered between the tip and the edge. We also
discuss possible explanations for spatial asymmetry of experimentally measured
resistance maps, and connect it with specific configurations of the measuring
probes.

\end{abstract}
\maketitle
\section{Introduction}

Up to now, scanning gate microscopy ({SGM}) has been successfully used to
image local electron transport in various mesoscopic systems. In early
applications, SGM probed the interference effects produced by microscopic
disorder in graphene,~\cite{berezovsky_sgm_1, berezovsky_sgm_2} and it was
later applied to image electron-hole puddles originating from extrinsic local
doping.~\cite{jalilian_sgm} In the later experiment,~\cite{jalilian_sgm} the
tip was coated with a dielectric and placed directly in contact with graphene.
The advantage of this approach is that {AFM} topography scans could be
performed simultaneously with {SGM} scans, and the tip could be additionally
used to clean the sample. Scanning technique was used in
Ref.~\onlinecite{connolly_sgm_1} to study the spatial inhomogeneity of the
local neutrality point, and to measure the efficiency of intentionally
embedding (writing) charges in graphene.~\cite{connolly_sgm_2} Tip-dependent
resistance map of a narrow quantum point contact (QPC), presented in
Ref.~\onlinecite{neubeck_sgm}, revealed a significant resistance increase when
the tip was placed directly above the sample. The technique is also suitable
for the investigation of localization effects. For example, concentric
conductance halos were observed in SGM maps scanned around localized states in
graphene quantum dots,~\cite{schenz_prb_sgm, schenz_njp_sgm} narrow
constrictions,~\cite{connolly_sgm_prb, garcia_sgm} and enhanced conductance
was reported in narrow nanoribbons.~\cite{pasher_sgm} Scanning gate
experiments in quantum point contacts were simulated in
Ref.~\onlinecite{mrenica_sgm}.


In this paper, we investigate the scanning gate experiments of the magnetic
focusing devices reported in Refs.~\onlinecite{bhandari_nanolett_2016,
bhandari_arxiv_2016, bhandari_new_arxiv, morikawa_sgm}. Before, similar
magnetic focusing measurements were performed on semiconductor two-dimensional
electron gas (2DEG) in parallel with the scanning technique.~\cite{aidala} We
model our device using both quantum and classical transport theory.
Previously, classical simulation of such a focusing device was done in
Ref.~\onlinecite{bhandari_nanolett_2016}, and focusing without SGM tip was
studied in Ref.~\onlinecite{slavisa_focusing}. A recent paper by
Kolasi{\'n}ski {\it et al.\/}~\cite{kolasinski_sgm_2} was the first to
reproduce some of the experimental findings by applying a full quantum
approach. Here, we implement the multi-terminal Landauer-B\"uttiker formalism
to calculate the non-local resistances. We additionally compare the resistance
maps of 4-terminal and 6-terminal devices for different combinations of
voltage probes. Due to the large size of the device, as compared to the
electron wavelength, most of the simulated effects can already be captured by
the computationally less demanding classical model. In respect to that, we
confirm that the classical billiard model can be successfully used to describe
transport of relativistic carriers in the presence of inhomogeneous
potentials, if the linear graphene spectrum is properly implemented. 


Our analysis of simulated SGM maps confirms and expands on some of the
experimental observations reported in
Refs.~\onlinecite{bhandari_nanolett_2016, bhandari_arxiv_2016,
bhandari_new_arxiv, morikawa_sgm}. We consider both positively and negatively
charged tip, as well as a tip acting in the mixed regime. We show that the
main resistance feature (e.g.~the spatial area of reduced resistance) can be
explained by considering geometric relations between the device boundaries,
the tip position, and the circular cyclotron orbits. We also found that the
finite width of the focusing leads can not be neglected. One of the novel
results is that the SGM tip is mainly acting on a set of trajectories that
directly connect the injector lead with the tip position. We show that the
repelling tip leaves a shadow behind itself, which is mainly situated in areas
delimited by two direct cyclotron orbits that connect the tip with the two
corners of the injector lead. The specific shape of the tip shadow, which
depends on the relative position between the tip and the injector lead,
determines the shape of the low resistance region. Therefore, at the first
focusing peak, the tip is imaging a specific subset of direct trajectories
connecting the two focusing leads. This only applies for a repelling tip, or a
tip operating in a mixed regime (repelling and focusing). For a tip acting as
a focusing lens, the produced SGM maps do not show any significant change in
the resistance, therefore they do not provide much useful information in
characterization of transport. We additionally compare results between a
4-terminal and a 6-terminal device, and find them to be very similar, although
the later induces some asymmetry in the resistance maps. The reason for this
similarity is because the resistance is mostly determined by the transmission
function between the two focusing leads. In that sense, we confirm the
approach taken in Ref.~\onlinecite{bhandari_nanolett_2016}, which considered
only transmission functions and not resistances.


This paper is organized as follows. In Sec.~\ref{sec_method} we describe the
focusing system, and indicate how we model the tip potential. Next, in
Sec.~\ref{sec_rmaps}, we scale the graphene tight-binding Hamiltonian and
compare relevant quantities, such as the dispersion relation and the current
density in a scaled and an unscaled graphene lattice. The resistance of an
unperturbed device (i.e~a device without the SGM tip) is studied in
Sec.~\ref{sec_sgm}, while the analysis of SGM scans is done in
Sec.~\ref{sec_sgm}. In Sec.~\ref{sec_6term} we present results of a
six-terminal device, and discuss the possible causes of spatial asymmetry seen
in the experiments. A short summary of our findings is given in
Sec.~\ref{sec_end}.

\section{Focusing system and SGM potential}\label{sec_method}

\begin{figure}[ht]
\includegraphics[width=0.48\textwidth]{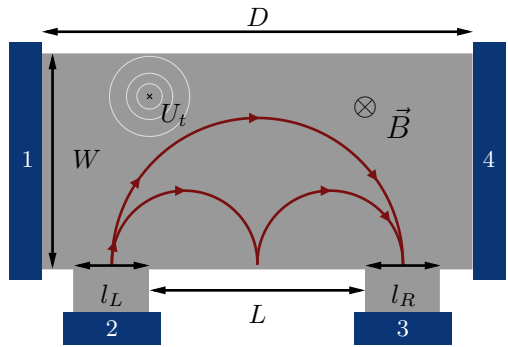}
\caption{\label{fig:system}
  Graphene magnetic focusing device: the system width is $W=2\,\mu\rm m$,
  while the system length is $D=4\,\mu\rm m$. Both horizontal (armchair), and
  vertical (zigzag) leads are metallic. The vertical leads have the same width
  $l_L=l_R=0.7\,\mu\rm m$, and their separation is $L=2\,\mu\textrm{m}$. The
  magnetic field is perpendicular to the graphene sheet and points in the
  negative $z$ direction. The potential profile of the {AFM} tip is
  schematically represented by the white circles.} 
\end{figure}

The studied system is shown in Fig.~\ref{fig:system}. It is a four-terminal
graphene device with the same dimensions as those used in
Ref.~\onlinecite{bhandari_nanolett_2016}. The only major difference between
our system, and that of Ref.~\onlinecite{bhandari_nanolett_2016}, is the
absence of two upper leads. We implement these two leads in
Sec.~\ref{sec_6term}, and discuss the changes they introduce in the SGM maps.

Magnetic focusing occurs when electron trajectories from the $2^{\rm nd}$ lead
(red curves in Fig.~\ref{fig:system}) are bent due to an external magnetic
field into the $3^{\rm rd}$ lead. Depending on the ratio between the width of
the $3^{\rm rd}$ lead ($l_R$) and the distance between the leads ($L$), the
diverted electron can exit into the $3^{\rm rd}$ or the $4^{\rm th}$ lead.
This switching of the exit lead manifests itself as oscillations in the device
resistance. Figure~\ref{fig:system} shows two focusing orbits, where an even
number of cyclotron radii matches the separation between the leads,
$2nR_c=\left[L + (l_R + l_L)/2 \right]$. Focusing occurs as long as $2R_c >
l_R$.

Although magnetic focusing is a local phenomena, which depends on specific
paths an electron can take in the system, it is usually studied by measuring
the resistance of the whole device. A lot of information is lost in such
measurements (e.g.~the most probable electron trajectories). This spatial
information can be retained by scanning-gate measurements, where a SGM tip
perturbs the circular electron trajectories, causing the device conductance to
become tip-dependent. The conductance maps produced in such measurements
reveal how device conductance depends locally on electron passage through that
point.


According to Refs.~\onlinecite{bhandari_arxiv_2016, bhandari_nanolett_2016}, a
charged {STM} tip placed above a graphene sheet modifies the local charge
density in graphene 
\begin{equation}
    \Delta n(\vec{r}) = \frac{\tilde{q} h}
                             {{\left(d {(\vec{r})}^2 + h^2\right)}^{3/2}}
\end{equation}
\noindent which depends on the tip relative charge $\tilde{q}=-q / 2\pi e$
(here $q$ is the actual charge accumulated on the tip, and $e$ is the electron
charge), the distance from the tip to the graphene plane $h$, and the distance
from the tip in-plane projection to the current point
\mbox{$\vec{d}(\vec{r})=\vec{r} - \vec{r}_{\textrm{tip}}$}. A local change in
the charge density modifies the local Fermi energy
\begin{equation}
    \Delta E_F(\vec{r})=E_F(n)-E_F(n+\Delta n(\vec{r})),
\end{equation}
\noindent which manifests as an additional tip-induced potential
\mbox{$U_t(\vec{r}) = \Delta E_F(\vec{r})$}. Since Fermi energy in graphene
depends on the charge density through \mbox{$E_F(n) = \hbar v_F\sqrt{\pi n}$},
the induced potential is
\begin{equation} U_t(\vec{r}) = \hbar v_F \sqrt{\pi} \left( \sqrt{n} - \sqrt{n
                + \Delta n(\vec{r})}\right). 
\end{equation}
\noindent This potential is a function of the global charge density $n$, and
the local charge modification $\Delta n(\vec{r})$. The global density $n$ is
set by the back-gate voltage, while the local modification $\Delta n(\vec{r})$
is determined by the tip height and the tip charge.


\begin{figure}[thb]
\includegraphics[width=0.48\textwidth]{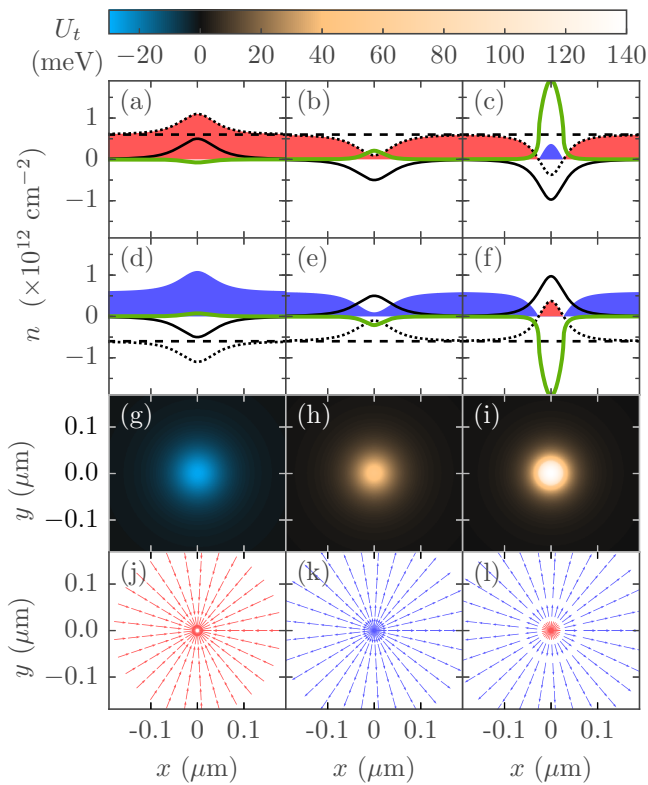}
\caption{\label{fig:tip_potential}
    Tip-induced potential: (a)--{(f)} Charge density beneath the tip
    ($x_{\rm tip}=y_{\rm tip}=0$). Solid black curves show the local
    charge modification $\Delta n(\vec{r})$, while dashed black
    lines mark the global (unperturbed) charge density $n$. Dotted black
    curves show the resulting density $\tilde{n}(r) = n + \Delta n(r)$.
    Colored areas present the absolute density $\left|\tilde{n}(r)\right|$
    for electrons (red) and holes (blue), while induced potential,
    expressed in charge density units $U_t^2/(\hbar^2 v_F^2\pi)$, is shown
    by the green curves. (g)--{(i)} Tip-induced potential for three
    regimes in (a),~(b),~and~(c), respectively. (j)--{(l)} Directions of
    the force field acting on the charge carriers: inward (red), and outward
    (blue), for the corresponding potentials in (g),~(h),~and~(i),
    respectively.}
\end{figure}


An alternative way to look into the perturbation of the tip is through the 
tip-generated force field. Charge carriers that travel through the system
experience a force $\vec{F}(\vec{r}) = -\vec{\nabla}U_t(\vec{r})$ generated
by the tip. This
force modifies the carrier equation of motion~\cite{bhandari_nanolett_2016}
\begin{equation}
\frac{d^2\vec{r}}{dt^2} = \frac {\vec{F}(\vec{r})} {m^{*}}
                = \frac{1}{2} v_F^2
                  \frac{\nabla \tilde{n}(\vec{r})}
                       {\tilde{n}(\vec{r})},
\end{equation}
\noindent where $m^{*}$ is the carrier dynamical mass in graphene
($m^{*}=\hbar \sqrt{\pi n}/v_{F}$), and
\mbox{$\tilde{n}(r) = n + \Delta n(r)$} is the resulting charge density.


Fig.~\ref{fig:tip_potential} shows tip-induced potentials for different
combinations of the global charge density and the local charge modification.
There are six different regimes in which the tip can operate, but only three
of these are unique. The other three regimes can be obtained by exchanging
electrons with holes. The first regime is presented in
Fig.~\ref{fig:tip_potential}{(a)}. Here, a positively charged tip increases
the local electron density, which manifests in the negative potential profile
shown in Fig.~\ref{fig:tip_potential}{(g)}. The tip-induced force field in
Fig.~\ref{fig:tip_potential}{(j)} reveals a focusing nature of the tip. The
case of negatively charged tip in Fig.~\ref{fig:tip_potential}{(b)} was
previously studied experimentally.\cite{bhandari_nanolett_2016} As shown in
Ref.~\onlinecite{bhandari_nanolett_2016} and in
Fig.~\ref{fig:tip_potential}{(h)}, the tip creates a positive potential which
then repels the incoming electrons. The force field in this regime, shown in
Fig.~\ref{fig:tip_potential}{(k)}, is pointing away from the tip. In both
Figs.~\ref{fig:tip_potential}{(a)} and~\ref{fig:tip_potential}{(b)}, we set
\mbox{$|\Delta n(\vec{r})|$} to $5 \times 10^{11}\,\textrm{cm}^{-2}$. This
density was used in Ref.~\onlinecite{bhandari_nanolett_2016} to fit the
experimental data, and it corresponds to a tip positioned \mbox{$h$ = 60 nm}
above the graphene sheet. Assuming that the tip charge $\tilde{q}$ does not
depend on the tip height $h$, in the far-left column of
Fig.~\ref{fig:tip_potential}, we present results for a tip positioned closer
to the sample (\mbox{$h$ = 43 nm}). When changing the tip height, we first
calculate the tip charge $\tilde{q}$ from the modified charge density
\mbox{$\Delta n(r_{\textrm{tip}}) = \tilde{q}/h^2$}, and then we recalculate
$\Delta n$ for the new height. 

Contrary to a classical 2DEG, where the tip depletes the electron density
beneath it,~\cite{szafran} in graphene, due to its gapless nature, the
depleted electrons turn into holes. The induced potential in this third (or
mixed) regime is much stronger than in both, the focusing and repelling
regimes (compare Fig.~\ref{fig:tip_potential}{(i)} with
Figs.~\ref{fig:tip_potential}{(g)} and~\ref{fig:tip_potential}{(h)}). The
force field is also specific (see Fig.~\ref{fig:tip_potential}{(l)}): in $n$
region (surrounding the tip) the tip repels incoming electrons, while in $p$
region (beneath the tip) the tip focuses tunneling holes. The actual tip
height used in the experiment~\cite{bhandari_nanolett_2016} could not go below
50 nm (because of the \mbox{$50$ nm} thick BN layer separating the tip and the
graphene sample). However, this does not mean that the third scanning regime
is experimentally inaccessible. Enhanced tip potentials can be realized by
lowering the global electron density $n$, or by increasing the tip charge
$\tilde{q}$.

As stated above, we model this system from two perspectives, the quantum and
the classical one. The classical billiard model is the same as that used in
Ref.~\onlinecite{slavisa_focusing}, while the quantum simulations are performed
using KWANT,~\cite{kwant} a software package for quantum transport.
Magnetic field in the quantum model is implemented through Peierls phase, 
as explained in Ref.~\onlinecite{petrovic_qhe}, and the resistances 
are obtained by applying the Landauer-B{\"u}ttiker formalism for a 
four-terminal~\cite{buttiker_4term} and 
multi-terminal~\cite{ferry} device.
\begin{figure}[t]
\includegraphics[width=0.48\textwidth]{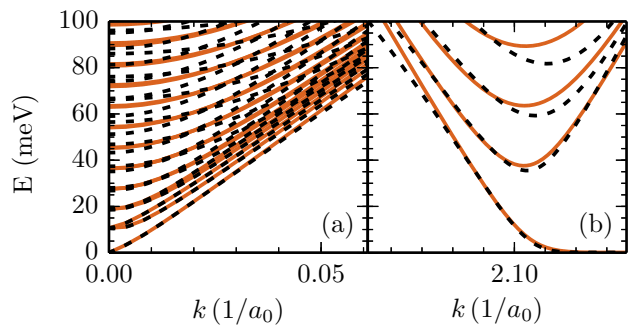}
\caption{\label{figDispersion}
    Dispersion relations of the $1^{\rm st}$ (a) and the $2^{\rm nd}$ (b) lead.
    Magnetic field is $B = 0.1\,\rm T$. Energy bands of pristine
    graphene are shown in orange, while those of the scaled
    graphene ($s_f = 15.15$) are shown by the black dashed curves. 
    System dimensions are ten times smaller than those used in
    Fig.~\ref{fig:system}. Since the scaling procedure modifies the inverse
    ($k$) space, we translated the K-point of the scaled system in (b), to
    match it with the K-point of the pristine graphene lattice.}
\end{figure}
\section{Scaling the tight-binding Hamiltonian}
In order to simulate devices of similar sizes as those used in the
experiment,\cite{bhandari_nanolett_2016} we scale the graphene tight-binding
Hamiltonian. As explained in Ref.~\onlinecite{liu_prl_2015}, a scaling
coefficient $s_f$ is introduced. This coefficient increases the spacing
between carbon atoms \mbox{$a = s_f a_0$}, and simultaneously decreases the
nearest neighbor hopping energy \mbox{$t = t_0 / s_f$}. The scaling procedure
allows for simulations of systems with dimensions comparable to those
used in actual experiments (in order of microns), but with lesser number of
tight-binding orbitals. However, the scaling has its limits. 
Results for
larger $s_f$ are less accurate, particularly for higher energies, away from
the linear part of the spectrum. Close to the Dirac point, the scaled system
is still a good approximation of the pristine graphene lattice.

To test the validity of the scaling procedure, we compare in
Fig.~\ref{figDispersion} the dispersion relations obtained using scaled
Hamiltonian with those obtained using the pristine graphene lattice. Since
performing tight-binding calculations on a micrometer scale is computationally
very demanding, we test the scaling procedure on a system ten times smaller
than that presented in Fig.~\ref{fig:system}. The lattice scale of this
smaller system \mbox{($s_f = 15.15$)} is comparable to the lattice scale we
use in the rest of the paper to simulate the micrometer-sized system shown in
Fig.~\ref{fig:system} (\mbox{$s_f = 15.34$}). As expected, the scaled
dispersions match the pristine graphene lattice dispersions for low energies
(below 100~meV). Although the band minimal energies differ, the scaled lattice
is a good approximation for states away from the subband minima. Note that the
scale is chosen such that it preserves the metallic nature of the armchair
leads, as is seen in Fig.~\ref{figDispersion}{(a)}.

\begin{figure}[t]
\includegraphics[width=0.48\textwidth]{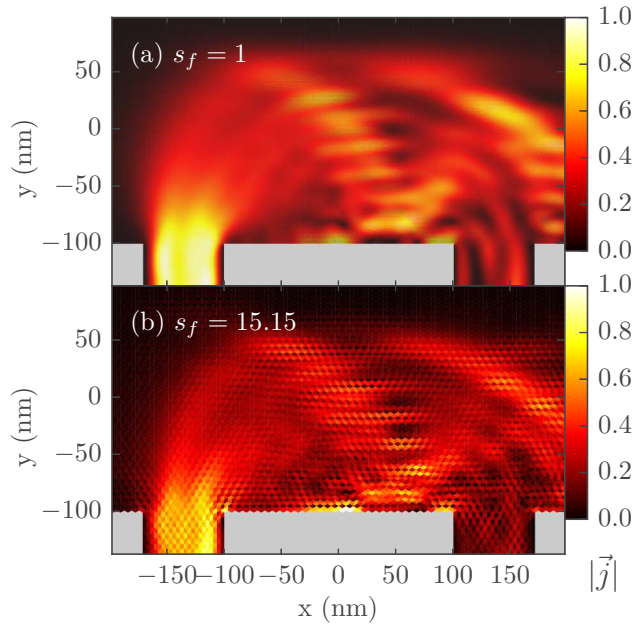}
\caption{\label{figScaledCurrent}
    (Color online) Normalized current density when current is injected from
    the second lead: (a) pristine graphene lattice ($s_f=1$), and (b) scaled
    graphene lattice
    ($s_f=15.15$). Magnetic field is $B=1\,\rm T$, and Fermi
    energy is $E_F=80\rm\,meV$. The presented system is ten times smaller
    than that in Fig.~\ref{fig:system}, SGM tip is not present.}
\end{figure}

Since SGM experiments probe the local properties, it is necessary to determine
how scaling affects them. With that in mind, in Fig.~\ref{figScaledCurrent} 
we compare two current densities: one for scaled, and one for 
unscaled lattice. Beside the loss in resolution, caused by a lesser number of
carbon atoms, the general current flow patterns are preserved with scaling,
which confirms that this method can be used to simulate SGM experiments.
\section{Magnetic focusing}\label{sec_rmaps}
\begin{figure}[htb]
\begin{center}
\includegraphics[width=0.48\textwidth]{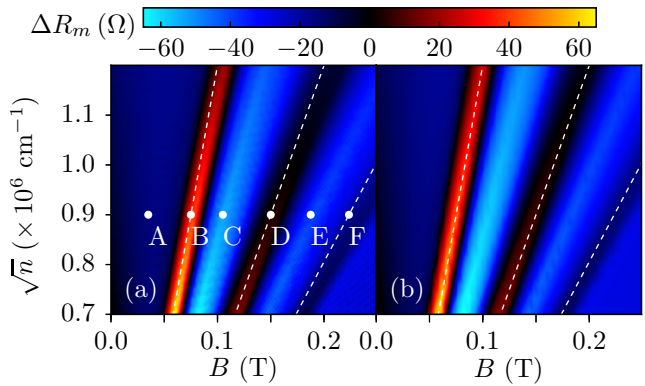}
\end{center}
\caption{\label{figRmaps}
   (Color online) Change of the $R_m = R_{12,43}$ resistance versus the 
   magnetic field and
   the electron density (Fermi energy) obtained from (a) quantum, and (b)
   classical simulation. In order to be able to compare both results with
   those published in Ref.~\onlinecite{bhandari_nanolett_2016}, we calculate
   $\Delta R_{m}$ by subtracting $R_{0} = 28\ \Omega$ from numerically
   obtained $R_{12,43}$. The white dashed lines mark the first three focusing
   maxima, and they correspond to cyclotron radii of $1.4\ \mu\rm m$,
   $0.7\ \mu\rm m$, and $0.47\ \mu\rm m$. Labels {A}--{F} mark the ($\sqrt{n},
   B$) points for which we present the current density in
   Fig.~\ref{fig_jfocus}.} 
\end{figure}

We investigate now the general transport properties of the focusing device
when no SGM tip is present. We are interested in how the focusing resistance
($R_m$ = $R_{12,43}$) changes as a function of the magnetic field $B$, and the
electron density $n$ (or Fermi energy).
Ref.~\onlinecite{bhandari_nanolett_2016} reported several resistance peaks as
a function of the applied field, related with the current focusing (see
Fig.~2(a) in Ref.~\onlinecite{bhandari_nanolett_2016}). We simulate these
measurements, and in Fig.~\ref{figRmaps} we present a comparison of the
resistances obtained from both, quantum and classical simulation. To match our
colormaps with those of Ref.~\onlinecite{bhandari_nanolett_2016}, in both
cases, we subtract $R_{0} = 28\ \Omega$ from numerically calculated $R_{12,
43}$. This value is very close to $(R_m^{\max} + R_m^{\min})/2$. The
resistances obtained with the two methods agree both qualitatively and
quantitatively. The only difference is that resistances obtained with the
quantum method show a set of parabolic fringe lines at higher fields, coming
from Landau quantization. The classical method does not account for
transversal quantization in the leads, and for the existence of transverse
modes, hence the transmissions obtained classically need to be properly scaled
before resistance calculations. We perform this scaling by multiplying the
classically obtained transmissions with the approximate number of modes in the
source lead~\cite{data}
\begin{equation}
    M_i = 2 \frac{E_F\, W_i}{\hbar\, v_F\, \pi}.
\end{equation}
\noindent Here index $i$ refers to the leads, $W_i$ is the width of the $i$-th
lead, $E_F$ is the Fermi energy in the $i$-th lead, $v_F$ is the graphene
Fermi velocity, while coefficient 2 is added to account for contributions
coming from the two valleys. The graphene Fermi velocity $v_F$ in the
tight-binding model is determined (to first approximation) from the
nearest-neighbour hopping energy as $v_F = 3|t|a/2\hbar$, where $a$ is the
distance between neighbouring carbon atoms. For $t = 2.7$~eV, and $a =
1.42$~\AA~we obtain $v_F = 873\,893$~m/s, which is the value that we use in
our classical model.


Since cyclotron radius in graphene is proportional to the Fermi energy $R_c =
E / (ev_F B)$, and Fermi energy is proportional to carrier density $E = \hbar
v_F \sqrt{\pi n}$, then $\sqrt{n} = \gamma R_c B$ (where $\gamma = e/(\hbar
\sqrt{\pi})$). In other words, for equiradial lines, $\sqrt{n}$ is a linear
function of the applied magnetic field. In Fig.~\ref{figRmaps} we mark three
such lines (white dashed lines) for three cyclotron radii (1.4 $\mu$m, 0.7
$\mu$m, and 0.47 $\mu$m) in order to match the first three focusing maxima.

\begin{figure}[tb]
\begin{center}
\includegraphics[width=0.48\textwidth]{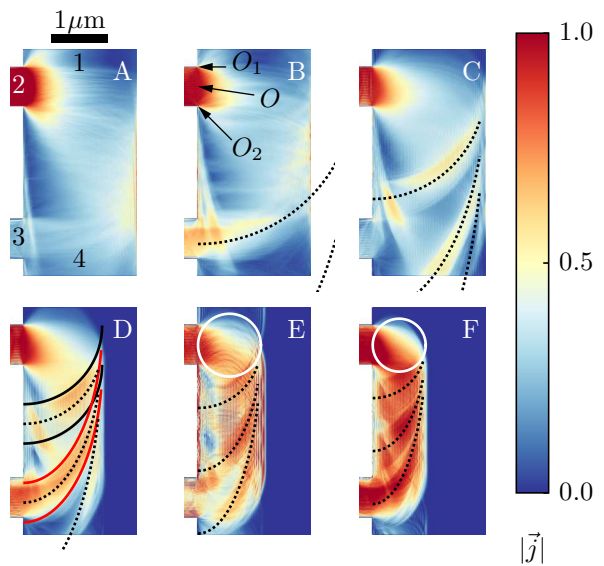}
\end{center}
\caption{\label{fig_jfocus}
   (Color online) Normalized total current density at different focusing
   fields: (A) $35\,\rm mT$,
   (B) $75\,\rm mT$, 
   (C) $105\,\rm mT$, 
   (D) $150\,\rm mT$, 
   (E) $188\,\rm mT$, and 
   (F) $223\,\rm mT$ 
   (points {A}--{F} marked in Fig.~\ref{figRmaps}). The current is calculated
   using the quantum method. The electron density is $n = 8.1 \times
   10^{11}\,{\rm cm}^{-2}$. Lead numbers for this rotated system are shown in
   figure A. Black-dashed curves show elliptical trajectories given by
   Eq.~(\ref{eq_elipsa}), and centered on the middle of the $2^{\rm nd}$ lead
   (point $O$). Black and red full curves show the same trajectories, but now
   centered at the corners of the $2^{\rm nd}$ lead (points $O_1$ and $O_2$). 
   White circles in E and F show cyclotron orbits for these two figures.}
\end{figure}

For narrow focusing leads (leads 2 and 3), resistance peaks appear each time a
multiple of a cyclotron diameter $2R_c$ matches the lead distance $L$.
However, if the lead widths ($l_R$ and $l_L$) are comparable to the distance
between them, it is reasonable to assume that focusing would occur only when
electron injected from the middle of the input lead exits in the middle of the
output lead. Two such trajectories are presented in Fig.~\ref{fig:system}. A
simplified focusing formula, which includes the lead widths, would then be
$2nR_c = (L + l)$. For $L = 2\ \mu{\rm m}$ and $l = 0.7\ \mu{\rm m}$, the
first three focusing radii are 1.35 $\mu$m, 0.675 $\mu$m and 0.45 $\mu$m,
which approximately matches the three lines shown in Fig.~\ref{figRmaps}. For
cyclotron radii smaller than half of the lead width (0.35 $\mu$m) focusing is
no longer possible.


To test whether the three resistance peaks in Fig.~\ref{figRmaps} appear due
to the current focusing, we show in Fig.~\ref{fig_jfocus} the local current
density for electrons coming from the $2^{\rm nd}$ lead, for points A-F marked
in Fig.~\ref{figRmaps}{(a)}. The current is obtained using the quantum
(tight-binding) model. Insets B, D, and F indeed show high current
concentration in the exiting ($3^{\rm rd}$) lead, which confirms our
assumption. For stronger fields (insets D, E, and F), the current flows close
to the system lower edge (the left edge in Fig.~\ref{fig_jfocus}), in an area
one cyclotron diameter wide, The reason why current spreads in an area one
cyclotron diameter wide, and not one cyclotron radius wide, is the following.
In the classical picture, electrons which enter the system perpendicularly to
the lower edge (i.e.~parallel to the focusing leads) would spread
approximately one cyclotron radius away from the edge, since their
trajectories consist of semicircles (see the two trajectories in
Fig.~\ref{fig:system}). On the other hand, electrons entering the system
almost parallel to the lower edge (i.e.~normal to the focusing lead direction)
would make almost a full circle before they scatter on the lower edge (see two
white circles in insets E and F in Fig.~\ref{fig_jfocus}). The current then
spreads in a diameter-wide area due to these electrons. For weak fields, most
of these electrons do not even make a full orbit, since they exit into the
$1^{\rm st}$ lead (see insets A and B in Fig.~\ref{fig_jfocus}, where one part
of the current from the $2^{\rm nd}$ lead exits into the $1^{\rm st}$ lead).

Current in the system can be understood in terms of the cyclotron orbits.
Based on the picture of classical trajectories presented in
Ref.~\onlinecite{bhandari_phd}, we plot three envelope curves (black-dotted
lines) to mark three paths where (according to the classical picture) the
current is supposed to travel. Each of these three curves is a part of an
elliptical line 
\begin{equation}
\label{eq_elipsa}
  {(x-x_{0})}^2 - {\left(\frac{y - y_0}{2n}\right)}^2 = R_c^2, 
  \quad\quad n = 1, 2, 3, \ldots
\end{equation}
\noindent centered on the middle of the input lead $(x_0, y_0) = O$ (see point
$O$ in Fig.~\ref{fig_jfocus}B). It is clear that focusing occurs in insets B,
D, and F, since for these insets the elliptic curves pass through the $3^{\rm
rd}$ lead. The input lead has a finite width, hence these trajectories spread
in space. The spread of current density is delimited by elliptical curves
defined by Eq.~(\ref{eq_elipsa}) (black and red full curves in
Fig.~\ref{fig_jfocus}D), but now centered on two corners of the input lead
(see points $O_1$ and $O_2$ in Fig.~\ref{fig_jfocus}B).



\begin{figure}[tb]
\begin{center}
\includegraphics[width=0.48\textwidth]{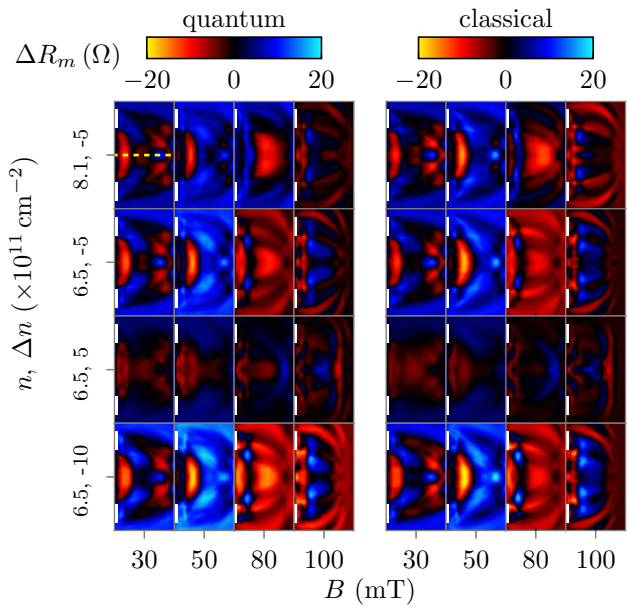}
\end{center}
\caption{\label{figSGM}
   (Color online) Comparison of the SGM resistance maps obtained from quantum
   (left) and classical (right) simulation. The two white stripes in the
   figures mark the positions of the $2^{\rm nd}$ and the $3^{\rm rd}$ lead.
   Labels on the left mark the electron density $n$, and maximal change in
   charge density introduced by the tip $\Delta n$. $\Delta R_m$ is calculated
   by subtracting $R_0 = (R_{12,43}^{\max} + R_{12,43}^{\min})/2$ from
   $R_{12,43}$ for each scan. Yellow-dashed line in the first row marks the
   mirror symmetry axis.} 
\end{figure}

\section{Scanning gate microscopy}\label{sec_sgm}

In this part we compare scanning gate maps obtained using the two models. We
mentioned above that there are three regimes in which the tip can operate,
therefore we additionally compare how scanning maps change in these three
regimes. This comparison is presented in Fig.~\ref{figSGM}. In general, the
local feature of the SGM maps obtained with the two models match. This is
expected since the system size is larger than the electron wavelength, hence
some of the interference effects are suppressed. The figure also confirms that
on these scales a computationally less demanding classical model manages to
capture all the features obtained with a more detailed atomistic model. 

The first two rows in Fig.~\ref{figSGM} show SGM maps for a repulsive
(negatively charged) tip, as it was used in the
experiment.~\cite{bhandari_nanolett_2016} The calculated resistances are very
similar to the measured ones,~\cite{bhandari_nanolett_2016} and the main
difference is that our results posses some extra oscillations close to the
upper edge (shown on the right side in the rotated system in
Fig.~\ref{figSGM}). As we show below, these features originate from multiple
electron scatterings between the tip and the upper edge. The defining
characteristic of all SGM maps obtained with repelling tip is a region of
suppressed resistance close to the lower edge. This region evolves as magnetic
field is increased and as shown in recent experiments, it can be connected
with the cyclotron radius.~\cite{bhandari_new_arxiv}


Resistance maps obtained with the focusing (positively charged) tip (third row
in Fig.~\ref{figSGM}) convey less information than those obtained with a
repelling tip. Although the tip causes some change in the resistance, some
other effects, e.g~temperature smearing, would even more degrade the obtained
resistance maps. Therefore a tip in the focusing regime is probably not the
best choice to probe electron transport. The third (mixed) regime seems to
produce the largest change in the resistance: $R$-maps obtained in this regime
show almost identical features as those obtained with a repelling tip, but the
sample response is much better due to a stronger repelling force.


\begin{figure}[tb]
\begin{center}
\includegraphics[width=0.48\textwidth]{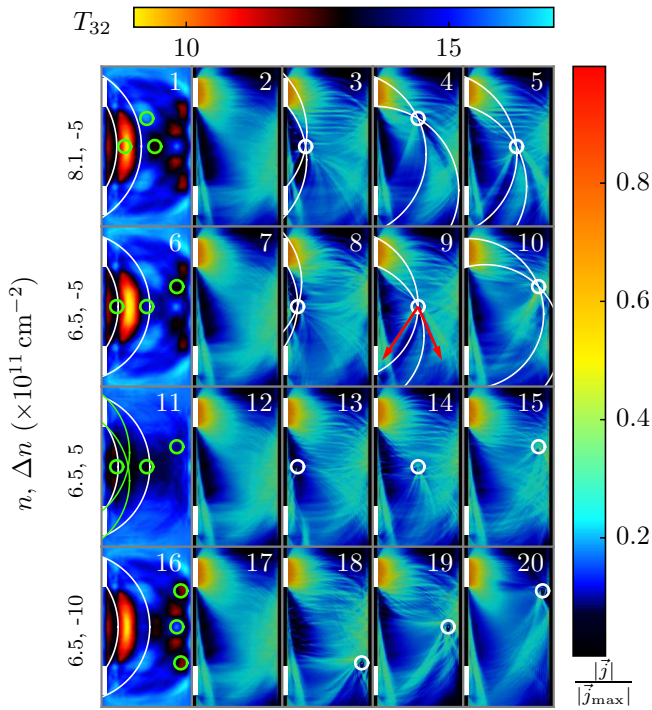}
\end{center}
\caption{\label{figT}
   (Color online) Comparison between electron transmission maps ($1^{\rm st}$
   column on the left) and normalized electron current densities (columns 2--5
   counted from the left). In the second column (insets 2, 7, 12, 17) we
   present current density with no SGM tip present, and in the rest (last
   three columns) we show current densities for specific tip positions (white
   circles). These tip positions are also marked in the first column (green
   circles) for comparison. The four rows correspond to the same four $n$,
   $\Delta n$ configurations as used in Fig.~\ref{figSGM}. Magnetic field is
   $B = 50$~mT. Again, the two focusing leads are marked in each inset with
   two white rectangles. In insets 3--5, and 8--10 we mark the direct
   trajectories connecting corners of the $2^{\rm nd}$ lead with the tip
   position. The corresponding cyclotron radii for the two group of lines are
   $2.1\ \mu$m (first row), and $1.88\ \mu$m (second row). The white lines in
   the first column delimit the areas relevant for focusing. } 

\end{figure}


Previous (classical) simulations of magnetic
focusing~\cite{bhandari_nanolett_2016} considered only electron transmissions
between the two focusing leads, and not the resistances. Here, we verify that
this approach is valid. In Fig.~\ref{figT} we compare the transmission maps
between the two focusing leads ($T_{32}$, left column) with the current
densities (columns 2--4) for some specific tip positions (white and green
circles). Calculated transmissions in the $1^{\rm st}$ column of
Fig.~\ref{figT} resemble the corresponding $R$-maps in Fig.~\ref{figSGM}
(columns for $B = 50$ mT in Fig.~\ref{figSGM}). Most of the features of the
resistance maps are determined by the tip position relative to the two corners
of the injector lead. In each of the insets 3--5, and 8--10, we show two
cyclotron orbits that directly connect the tip with the two corners (two white
arcs in each of these insets). As shown in these insets, a trail of suppressed
current is seen as a shadow that the tip leaves behind itself. The shadow is
situated mostly in the area enclosed between the two cyclotron orbits. These
two orbits (originating at two corners of the $2^{\rm nd}$ lead) mark the
boundary of a set of direct cyclotron trajectories that connect the tip with
the $2^{\rm nd}$ lead. The current is suppressed in these areas, because the
tip blocks these trajectories. The diverted current forms an arc around the
tip, and it flows away from the blocked area (see the two red arrows in inset
9, showing the flow direction of the diverted current). Similar explanation
for the diverted current is given in Ref.~\onlinecite{morikawa_sgm}. Our
interpretation (based on the two delimiting orbits) also explains the areas of
strongly suppressed resistance. Insets 7 and 8 show two positions of the tip
lying on the edge of the resistance suppressed region (see inset 6). For any
point lying on a line between these two points, the trail of blocked current
coincides with the $3^{\rm rd}$ lead. From these simple geometric relations,
we see that the area of suppressed conductance is delimited by two cyclotron
orbits that directly connect inner and outer corners of the focusing leads
(white curves in the $1^{\rm st}$ column of Fig.~\ref{figT}). The maximal
resistance suppression is expected at the crossing point of orbits connecting
inner and outer corners (see the two green curves in inset 11, and also the
two cyclotron orbits in inset 3). Based on the previous analysis, we conclude
that the finite width of the suppressed resistance region is an indirect
consequence of the finite width of the two focusing leads. 


The results for a focusing tip reveal effects opposite to those of the
repelling tip (compare the tip influence on the current densities in the
second and the third row in Fig.~\ref{figT}). The focusing tip leaves a trail
of enhanced current instead of a shadow, as seen in the case of a repelling
tip (compare insets 10 and 15). In the last row, a tip in the mixed regime
shows a much darker shadow behind itself as compared to the repelling tip. The
mixed nature of the tip manifests itself in a current profile, where some of
the current that manages to tunnel through the potential induced by the tip,
exits focused on the other side.

\section{Resistance maps of a six-terminal device}\label{sec_6term}

Although previous resistance maps capture the main features reported in the
experiment~\cite{bhandari_nanolett_2016} (e.g.~they show semicircular areas
where the resistance is reduced, and these areas coincide with the focusing
trajectories), they posses some additional features which were not observed
experimentally. The major difference is that the simulated resistance maps are
symmetric with respect to mirror reflection along the middle line of the
system (the yellow-dashed line in the first row of Fig.~\ref{figSGM}). The
cyclotron orbits imaged in the experiment~\cite{bhandari_nanolett_2016} were
not perfectly symmetric with respect to this transformation. This asymmetry
could originate from several different sources. For example, it could come
from local impurity charges trapped in the sample. Due to the electric forces
coming from these charges, transmitted electrons could divert from their ideal
circular trajectories. Although this is a possible explanations for the
asymmetry, it is unlikely in samples sandwiched between h-BN, hence here we
will discuss other possible sources. 


Our initial assumption was that the asymmetry could originate from a
difference in widths of the two focusing leads (the $2^{\rm nd}$ and the
$3^{\rm rd}$ lead). We tested this by changing the width of the $3^{\rm rd}$
lead, and recalculating some of the resistance maps of Fig.~\ref{figSGM}.
Using a wider ($l = 1.2\,l_{{\rm L}}$), or a narrower ($l = 0.8\,l_{{\rm L}}$)
$3^{\rm rd}$ lead did not significantly change the symmetry of the resistance
maps and can not account for what is observed in the experiment. Our second
assumption was that the asymmetry originates from an asymmetry in the
tip-induced potential. An uneven distribution of charges on top of the tip, or
a tip not properly aligned to the vertical ($z$) axis would create an
anisotropic image-charge density, and consequently an anisotropic tip
potential. We tested this by modifying the eccentricity of an elliptic charge
density, but the obtained resistance maps were not significantly modified. In
general, we could not reproduce the measured resistance asymmetries in an
impurity-free four-terminal device.


Since the original experiment was performed in a six-terminal device, in order
to check how resistance maps change for different configurations of the
voltage measuring probes, we additionally add two new leads to our system. We
label these new leads as lead 5, and lead 6, and they are placed opposite to
the two focusing leads. Our system is now a symmetric, six-terminal Hall bar
often used in standard quantum Hall measurements. 


\begin{figure}[tb]
\begin{center}
\includegraphics[width=0.49\textwidth]{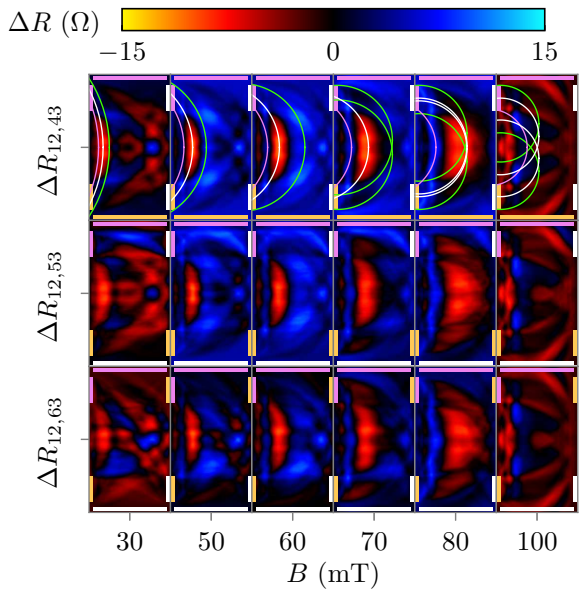}
\end{center}
\caption{\label{fig_6term}
   (Color online) SGM resistance maps of a six-terminal device for different
   measurement configurations: $\Delta R_{12,43}$ (first row), $\Delta
   R_{12,53}$ (second row), $\Delta R_{12,63}$ (third row), and for different
   magnetic fields (columns). Leads colored in violet are source and drain
   leads for the current, and leads colored in orange are voltage probes
   (leads colored in white were not used in the calculations). Global electron
   density is $n = 8.1\times10^{11}\ {\rm cm}^{-2}$, and maximal tip-induced
   change in the charge density is $\Delta n = 5\times10^{11}\ {\rm cm}^{-2}$.
   All resistances are obtained using the quantum (tight-binding) approach.
   Resistance $R_0$ is calculated and subtracted from every inset,
   similarly as in Fig.~\ref{figSGM}.}
\end{figure}


Resistance maps of this system are shown in Fig.~\ref{fig_6term}, for three
different cases. The first row presents data for the same probe configuration
as in Fig.~\ref{figSGM}, but now with two new leads included. When compared
with Fig.~\ref{figSGM}, the new leads do not significantly modify the
calculated resistances. The region of suppressed resistance is still delimited
with two cyclotron orbits connecting the outer and inner corners of the two
focusing leads (violet and green curves in the first row of
Fig.~\ref{fig_6term}), and maximal resistance suppression is still determined
by the orbits connecting the middle lines of the two focusing leads (the white
curves). The second row in Fig.~\ref{fig_6term} corresponds to probe
configuration actually used in the experiment.~\cite{bhandari_nanolett_2016}
A slight asymmetry is introduced by keeping the current probes (leads 1 and
2), and changing the voltage probes (from measuring $V_{43}$, to measuring
$V_{53}$). Although the suppressed region is still fairly symmetric, an
asymmetry is evident if we compare upper and lower parts of the $R$-maps. The
third row shows $R$-maps obtained by measuring the voltage across the device
($V_{63}$). A clear asymmetry is evident for lower fields. In general, the
measured voltage depends mostly on how much of the electron current is
scattered into the voltage leads. For stronger fields, most of the current is
located away from the new leads, on the lower edge, and therefore the
scattering is negligible.


Since we demonstrated that most of the transport in this system is determined 
by considering electron orbits that connect the $2^{\rm nd}$ lead with the 
tip, the interface between the focusing leads and the main region might also 
play a significant role. Here, we considered the focusing leads connected 
to the main region with perfect (90$^\circ$ degrees) corners, but due 
to an imperfect etching, these corners might be more smooth, thus allowing 
for some additional effects (e.g.~new direct trajectories from the 
$2^{\rm nd}$ lead to the tip).

\label{sec_six_terminal}
\section{Summary}\label{sec_end}

To summarize, we performed simulations of the scanning gate measurements in
graphene magnetic focusing devices. Two methods (quantum and classical) were
used to obtain the system transmissions. These transmissions were then applied
in the (multi-terminal) Landauer-B{\"u}ttiker formula to calculate the device
resistances. In order to perform the quantum simulations, the graphene
tight-binding Hamiltonian needed to be properly scaled. 

In case without the SGM tip, the focusing resistance $R_{m}(E, B)$ reveals
three focusing peaks, which could be connected with three cyclotron radii.
These radii were calculated using the distance between the two focusing
leads, but only after including the finite widths of the focusing leads. 

Depending on the voltage on the back-gate and the charge accumulated on the
tip, we differentiate between six different regimes in which the tip can
operate. Three out of these six regimes are unique. In general, due to the
large system size, all the features of the resistance maps are captured with
the classical model, and can be explained by tip influencing the direct
cyclotron orbits coming from the $2^{\rm nd}$ lead. Our results show that
the largest change in the resistance is obtained for tip operating in the
mixed regime (simultaneously repelling and focusing electrons). The spatial
asymmetry in experimentally obtained $R$-maps can be partially explained by
the specific configuration of the voltage probes, but we do not rule out other
sources, such as charged impurities or edge imperfections produced
during the etching process. 

\section{Acknowledgement}
This work was supported by the Methusalem program of the Flemish government.

%

\end{document}